\documentclass{llncs}

\usepackage{amsmath}
\usepackage{amssymb}
\usepackage{mathtools}
\usepackage{xspace}
\usepackage{tikz}
\usepackage{fixme}
\usepackage{wrapfig}
\usepackage{paralist,mdwlist}
\usepackage{manfnt,mparhack}
\usepackage[colorlinks=true, citecolor=black,linkcolor=black,urlcolor=black]{hyperref}
\usepackage{multirow}
\usepackage{pdfsync}

\usepackage{macros}
\title{Non-Zero Sum Games for Reactive Synthesis\thanks{Work supported by the ERC starting grant {\sc inVEST} (FP7-279499), G.A.~P\'{e}rez is supported by F.R.S.-FNRS ASP fellowship, M.~Randour is a F.R.S.-FNRS Postdoctoral Researcher.}}

\author{Romain Brenguier\inst{1}, Lorenzo Clemente\inst{2}, Paul Hunter\inst{3},
Guillermo A. P\'{e}rez\inst{3},
	Mickael Randour\inst{3}, Jean-Fran\c{c}ois Raskin\inst{3}, Ocan Sankur\inst{4}, Mathieu Sassolas\inst{5}}

\institute{
University of Oxford, UK
\and
University of Warsaw, Poland
\and
Universit\'e Libre de Bruxelles, Belgium
\and
CNRS, Irisa, France
\and
Universit\'e Paris-Est\,--\,Cr\'eteil, LACL, France
}

\begin{document}

\maketitle

\begin{abstract}
In this invited contribution~\cite{lata16}, we summarize new solution concepts useful for the
synthesis of reactive systems that we have introduced in several recent publications. 
These solution concepts are developed in the
context of non-zero sum games played on graphs. They are part of the
contributions obtained in the {\sc inVEST} project funded by the European
Research Council.
\end{abstract}

\section{Introduction}

{\em Reactive systems} are computer systems that maintain a continuous interaction with the environment in which they operate. They usually exhibit characteristics, like real-time constraints, concurrency, parallelism, etc., that make them difficult to develop correctly. Therefore, formal techniques using mathematical models have been advocated to help to their systematic design.

One well-studied formal technique is {\em model checking}~\cite{ClarkeE81,QueilleS82,BK08} which compares a model of a system with its specification. The main objective of this technique is to find design errors early in the development cycle. So model-checking can be considered as a sophisticated {\em debugging} method. A scientifically more challenging goal, called {\em synthesis}, is to design algorithms that, given a specification for a reactive system and a model of its environment, directly synthesize a correct system, i.e., a system that enforces the specification {\em no matter how} the environment behaves. 

Synthesis can take different forms: from computing optimal values of parameters to the full-blown automatic synthesis of finite-state machine descriptions for components of the reactive system. The main mathematical models proposed for the synthesis problem are based on {\em two-player zero-sum games played on graphs} and the main solution concept for those games is the notion of {\em winning strategy}. This model encompasses the situation where a {\em monolithic} controller has to be designed to interact with a {\em monolithic} environment that is supposed to be {\em fully antagonistic}. In the sequel, we call the two players \eve\/ and \adam, \eve\/ plays the role of the system and \adam\/ plays the role of the environment.

A fully antagonistic environment is most often a {\em bold abstraction} of
reality: the environment usually has its own goal which, in general, does not
correspond to that of falsifying
the specification of the reactive system.  Nevertheless, this
abstraction is popular because it is simple and sound: a winning strategy
against an antagonistic environment is winning against any environment that
pursues its own objective. However this approach may fail to find a winning
strategy even if solutions exist when the objective of the environment are taken into account, or it may produce sub-optimal solutions because they are overcautious and do not exploit the fact the the environment has its own objective. In several recent works, we have introduced new
solution concepts for synthesis of reactive systems that take the objective of the
environment  into account or relax the fully adversarial assumption.

\paragraph{{\bf Assume admissible synthesis}}
In~\cite{BRS-concur15}, we proposed a novel notion of synthesis where the objective of the environment can be captured using the concept of {\em admissible} strategies~\cite{adam2008admissibility,berwanger07,BRS14}. 
For a player with objective $\phi$, a strategy $\sigma$ is {\em dominated} by $\sigma'$ if $\sigma'$ does as well as $\sigma$ w.r.t. $\phi$ against all strategies of the other players, and better for some of those strategies. A strategy $\sigma$ is {\em admissible} if it is {\em not} dominated by another strategy. 
We use this notion to derive a meaningful notion to \emph{synthesize} systems with several players, with the following idea.
Only admissible strategies should be played by {\em rational} players as dominated strategies are clearly {\em sub-optimal options}. 
In {\em assume-admissible synthesis}, we make the assumption that both players play admissible strategies. 
Then, when synthesizing a controller, we search for an admissible strategy that is \emph{winning} against all admissible strategies of the environment. 
Assume admissible synthesis is {\em sound}: if both players choose strategies that are winning against admissible strategies of the other player, the objectives of both players will be satisfied. 

\paragraph{{\bf Regret minimization: best-responses as yardstick}}
In~\cite{hpr15} we studied strategies for \eve\/ which \emph{minimize her
regret}. The regret of a strategy $\sigma$ of \eve\/ corresponds to the
difference between the value \eve\/ achieves by playing $\sigma$ against \adam\/
and the value she could have ensured if she had known the strategy of \adam\/ in
advance. Regret is not a novel concept in game theory see, e.g.,~\cite{hp12},
but it was not explicitly used for games played on graphs before~\cite{fgr10}.
The complexity of
deciding whether a regret-minimizing strategy for \eve\/ exists, and the memory
requirements for such strategies change depending on what type of behavior
\adam\/ can use. We have focused on three particular cases: arbitrary behaviors,
positional behaviors, and time-dependent behaviors (otherwise known as
\emph{oblivious} environments). The latter class of regret games was shown
in~\cite{hpr15} to be related to the problem of determining whether an automaton
has a certain form of determinism.

\paragraph{{\bf Games with an expected adversary}} In~\cite{bruyere_STACS2014,DBLP:journals/corr/BruyereFRR14,DBLP:conf/lics/ClementeR15}, we combined the classical formalism of two-player zero-sum games (where the environment is considered to be completely antagonistic) with \textit{Markov decision processes} (MDPs), a well-known model for decision-making inside a stochastic environment.
The motivation is that one has often a good idea of the \textit{expected behavior} (i.e., average-case) of the environment represented as a stochastic model based on statistical data such as the frequency of requests for a computer server, the average traffic in a town, etc. In this case, it makes sense to look for strategies that will maximize the \textit{expected performance} of the system. This is the traditional approach for MDPs, but it gives no guarantee at all if the environment deviates from its expected behavior,
which can happen, for example, if events with small probability happen,
or if the statistical data upon which probabilities are estimated is noisy or unreliable. On the other hand, two-player zero-sum games lead to strategies guaranteeing a \textit{worst-case performance} no matter how the environment behaves\,---\,however such strategies may be far from optimal against the expected behavior of the environment. With our new framework of \textit{beyond worst-case} synthesis, we provide formal grounds to synthesize strategies that \textit{both} guarantee some minimal performance against any adversary \textit{and} provide an higher expected performance against a given expected behavior of the environment\,---\,thus essentially combining the two traditional standpoints from games and MDPs.

\paragraph{{\bf Structure of the paper}}
Section~2 recalls preliminaries about games played on graphs while Section~3 recalls the classical setting of zero-sum two player games.
Section~4 summarizes our recent works on the use of the notion of admissibility for synthesis of reactive systems.
Section~5 summarizes our recent results on regret minimization for reactive synthesis.
Section~6 summarizes our recent contributions on the synthesis of strategies that ensure good expected performance together with guarantees against their worst-case behaviors.

\section{Preliminaries}

We consider two-player turn-based games played on finite (weighted) graphs. Such games are played on so-called weighted game arenas.

\begin{definition}[Weighted Game Arena]
A (turn-based) two-player {\em weighted game arena} is a tuple $\Ar=\warena$ where:
  \begin{itemize}
  	\item $\StEve$ is the finite set of states owned by \eve, $\StAdam$ is the finite set of states owned by \adam, $\StEve \cap \StAdam=\emptyset$ and we denote $\StEve \cup \StAdam$ by $\St$.
	\item $\Edges \subseteq S \times S$ is a set of edges, we say that $\Edges$ is {\em total} whenever for all states $\st \in \St$, there exists  $\st' \in \St$ such that $(\st,\st') \in \Edges$ (we often assume this w.l.o.g.).
	\item $\StInit \in \St$ is the initial state.
	\item $\wgt : \Edges \rightarrow \integer$ is the weight function that assigns an integer weight to each edge.
  \end{itemize}
\noindent 
We do not always use the weight function defined on the edges of the weighted game arena and in these cases we simply omit it.
\end{definition}
\noindent
Unless otherwise stated, we consider for the rest of the paper a fixed weighted game arena $\Ar=\warena$.

A {\em play} in the arena $\Ar$ is an {\em infinite} sequence of states $\pi=\st_0 \st_1 \dots \st_n \dots$ such that for all $i \geq 0$, $(\st_i,\st_{i+1}) \in \Edges$. 
A play $\pi=\st_0 \st_1 \dots$ is {\em initial} when $\st_0=\StInit$.
We denote by $\Plays(\Ar)$ the set of plays in the arena $\Ar$, and by $\InitPlays(\Ar)$ its subset of initial plays.

A {\em history} $\rho$ is a finite sequence of states which is a {\em prefix} of a play in $\Ar$. We denote by $\Pref(\Ar)$ the set of histories in $\Ar$, and the set of prefixes of initial plays is denoted by $\InitPref(\Ar)$. 
Given an infinite sequence of states $\pi$, and two finite sequences of states $\rho_1,\rho_2$, we write $\rho_1 < \pi$ if $\rho_1$ is a prefix of $\pi$, and $\rho_2 \leq \rho_1$ if $\rho_2$ is a prefix of $\rho_1$. 
For a history $\rho=\st_0 \st_1 \dots \st_n$, we denote by $\last(\rho)$ its last state $\st_n$, and for all $i,j$, $0 \leq i \leq j \leq n$, by $\rho(i..j)$ the infix of $\rho$ between position $i$ and position $j$, i.e., $\rho(i..j)=\st_i \st_{i+1} \dots \st_j$, and by $\rho(i)$ the position $i$ of $\rho$, i.e., $\rho(i)=\st_i$.
The set of histories that belong to \eve, noted $\PrefEve(\Ar)$ is the subset of histories $\rho \in \Pref(\Ar)$ such that $\last(\rho) \in \StEve$, and the set of histories that belong to \adam, noted $\PrefAdam(\Ar)$ is the subset of histories $\rho \in \Pref(\Ar)$ such that $\last(\rho) \in \StAdam$.

\begin{definition}[Strategy]
A {\em strategy} for \eve\/ in the arena $\Ar$ is a function $\StrEve : \PrefEve(\Ar) \rightarrow \St$ such that for all $\rho \in \PrefEve(\Ar)$, $(\last(\rho),\StrEve(\rho)) \in \Edges$, i.e., it assigns to each history of $\Ar$ that belongs to \eve\/ a state which is a $\Edges$-successor of the last state of the history. Symmetrically, a {\em strategy} for \adam\/ in the arena $\Ar$ is a function $\StrAdam : \PrefAdam(\Ar) \rightarrow \St$ such that for all $\rho \in \PrefAdam(\Ar)$, $(\last(\rho),\StrAdam(\rho)) \in \Edges$. The set of strategies for \eve\/ is denoted by $\Sigma_{\subeve}$ and the set of strategies of \adam\/ by~$\Sigma_{\subadam}$.
\end{definition}

When we want to refer to a strategy of \eve\/ or \adam, we write it $\sigma$. We denote by $\dom(\sigma)$ the domain of definition of the strategy $\sigma$, i.e., for all strategies $\sigma$ of \eve\/ (resp. \adam), $\dom(\sigma)=\PrefEve(\Ar)$ (resp. $\dom(\sigma)=\PrefAdam(\Ar)$).

A play $\pi=\st_0 \st_1 \dots \st_n \dots$ is {\em compatible} with a strategy $\sigma$ if for all $i \geq 0$ such that $\pi(0..i) \in \dom(\sigma)$, we have that $\st_{i+1}=\sigma(\rho(0..i))$. We denote by $\outcome_{\st}(\sigma)$ the set of plays that start in $\st$ and are compatible with the strategy $\sigma$. Given a strategy $\StrEve$ for \eve\/ and a strategy $\StrAdam$ for \adam, and a state $\st$, we write $\outcome_{\st}(\StrEve,\StrAdam)$ the unique play that starts in $\st$ and which is compatible both with $\StrEve$ and $\StrAdam$.

A strategy $\sigma$ is {\em memoryless} when for all histories $\rho_1,\rho_2 \in \dom(\sigma)$, if we have that $\last(\rho_1)=\last(\rho_2)$ then $\sigma(\rho_1)=\sigma(\rho_2)$, i.e., memoryless strategies only depend on the last state of the history and so they can be seen as (partial) functions from $\St$ to $\St$. 
$\Sigma^{\sf ML}_{\subeve}$ and $\Sigma^{\sf ML}_{\subadam}$ denotes memoryless strategies of \eve\/ and of \adam, respectively.
A strategy $\sigma$ is {\em finite-memory} if there exists an equivalence relation $\sim \subseteq \dom(\sigma) \times \dom(\sigma)$ of {\em finite index} such that for all histories $\rho_1,\rho_2$ such that $\rho_1 \sim \rho_2$, we have that $\sigma(\rho_1)=\sigma(\rho_2)$. If the relation $\sim$ is {\em regular} (computable by a finite state machine) then the finite memory strategy can be modeled by a finite state transducer (a so-called {\em Moore} or {\em Mealy} machine). If a strategy is encoded by a machine with $m$ states, we say that it has {\em memory size}~$m$.

An {\em objective} $\WinObj \subseteq \Plays(\Ar)$ is a subset of plays. A strategy  $\sigma$ is winning from state $\st$ if $\outcome_{\st}(\sigma) \subseteq \WinObj$.
We will consider both {\em qualitative objectives}, that do not depend on the weight function of the game arena, and {\em quantitative objectives} that depend on the weight function of the game arena.

Our qualitative objectives are defined with Muller conditions (which are a canonical way to represent all the regular sets of plays). Let $\pi \in S^{\omega}$, be a play, then
$\infocc(\pi)=\{ \st \in \St \mid \forall i \cdot \exists j \geq i \geq 0 : \pi(j)=\st \}$ is the subset of elements of $\St$ that occur infinitely often along $\pi$.
A Muller objective for a game arena $\Ar$ is a defined by a set of sets of states ${\cal F}$ and contains the plays $\{ \pi \in \St^{\omega} \mid \infocc(\pi) \in {\cal F} \}$. We sometimes take the liberty to define such regular sets using standard LTL syntax. For a formal definition of the syntax and semantics of LTL, we refer the interested reader to~\cite{BK08}.

We associate, to each play $\pi$, an infinite sequence of weights, denoted $\wgt(\pi)$, and defined as follows:

$$\wgt(\pi)=\wgt(\pi(0),\pi(1)) \wgt(\pi(1),\pi(2)) \dots \wgt(\pi(i),\pi(i+1)) \dots \in \integer^{\omega}.$$

To assign a value $\val(\pi)$ to a play $\pi$, we classically use functions like ${\sf sup}$ (that returns the supremum of the values along the play), ${\sf inf}$ (that returns the infimum), ${\sf lim sup}$ (that returns the limit superior), ${\sf lim inf}$ (that returns the limit inferior), ${\sf MP}$ (that returns the limit of the average of the weights along the play), or ${\sf dSum}$  (that returns the discounted sum of the weights along the play). We only define the mean-payoff measure formally.

Let $\rho=\st_0 \st_1 \dots \st_n$ be s.t. $(\st_i,\st_{i+1}) \in \Edges$ for all $i$, $0 \leq i < n$, the mean-payoff of this sequence of edges is $$\MP(\rho)=\frac{1}{n} \cdot \sum_{i=0}^{i=n-1} \wgt(\rho(i),\rho(i+1)),$$

\noindent
i.e., the mean-value of the weights of the edges traversed by the finite sequence $\rho$. 
The {\em mean-payoff} of an (infinite) play $\pi$, denoted $\MP(\pi)$, is a real number defined  from the sequence of weights $\wgt(\pi)$ as follows: $$\MP(\pi)=\liminf_{n \rightarrow +\infty} \frac{1}{n} \cdot \sum_{i=0}^{i=n-1} \wgt(\pi(i),\pi(i+1)),$$

\noindent
i.e., $\MP(\pi)$ is the limit inferior of running averages of weights seen along the play $\pi$. Note that we need to use $\liminf$ because the value of the running averages of weights may oscillate along $\pi$, and so the limit is not guaranteed to exist.

A game is defined by a (weighted) game arena, and objectives for \eve\/ and \adam.

\begin{definition}[Game]
A game ${\sf G}=(\Ar,\WinObj_{\subeve},\WinObj_{\subadam})$ is defined by a game arena $\Ar$, an objective $\WinObj_{\subeve}$ for \eve, and an objective $\WinObj_{\subadam}$ for \adam.
\end{definition}
	
\section{Classical Zero-Sum Setting}

In zero sum games, players have antagonistic objectives.

\begin{definition}
A game ${\sf G}=(\Ar,\WinObj_{\subeve},\WinObj_{\subadam})$ is {\em zero-sum} if $\WinObj_{\subadam}=\Plays \setminus \WinObj_{\subeve}$.
\end{definition}

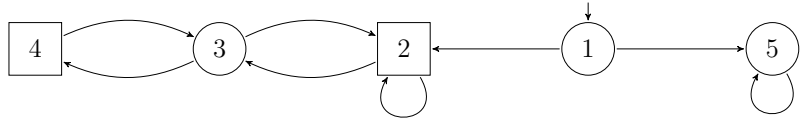
\begin{figure}[thb]
 \begin{center}
  \scalebox{0.7}{
\begin{tikzpicture}[->,>=stealth',shorten >=1pt,shorten <=1pt,auto,node
    distance=3.5cm,bend angle=45,font=\Large]
    \tikzstyle{p1}=[draw,circle,text centered,minimum size=10mm]
    \tikzstyle{p2}=[draw,rectangle,text centered,minimum size=10mm]
\node[p1,initial,initial where=above,initial text=] (q1) {$1$};
\node[p2,left of=q1] (q2) {$2$};
\node[p1,left of=q2] (q3) {$3$};
\node[p2,left of=q3] (q4) {$4$};
\node[p1,right of=q1] (q5) {$5$};

\path[->] (q1) edge (q2);
\path[->] (q1) edge (q5);
\path[->] (q2) edge[loop below,out=-60,in=-120,looseness=2, distance=12mm] (q2);
\path[->] (q2) edge[bend left=25] (q3);
\path[->] (q3) edge[bend left=25] (q2);
\path[->] (q3) edge[bend left=25] (q4);
\path[->] (q4) edge[bend left=25] (q3);
\path[->] (q5) edge[loop below,out=-60,in=-120,looseness=2, distance=12mm] (q5);

\end{tikzpicture}}
\vspace{-6mm}
  \end{center}
\caption{An example of a two-player game arena. Rounded positions belong to \eve, and squared positions belong to \adam.\label{fig:ex-adm}}
\end{figure}

\begin{example}
Let us consider the example of Fig.~\ref{fig:ex-adm}. Assume that the objective of \eve\/ is to visit $4$ infinitely often, i.e., $\WinObj_{\subeve}=\{ \pi \in \Plays \mid \pi \models \Box \Diamond 4 \}$, and that the objective of \adam\/ is $\WinObj_{\subadam}=\Plays \setminus \WinObj_{\subeve}$. Then it should be clear that \eve\/ does not have a strategy that enforces a play in $\WinObj_{\subeve}$ no matter what \adam\/ plays. Indeed, if \adam\/ always chooses to stay at state $2$, there is no way for \eve\/ to visit $4$ at all.
\end{example}

As we already said, zero-sum games are usually a bold abstraction of reality. This is because the system to synthesize usually interacts with an environment that has its own objective, and this objective is not necessarily the complement of the objective of the system. A classical way to handle this situation (see e.g.,~\cite{BloemEJK14}) is to ask the system to win only when the environment meets its own objective.

\begin{definition}[{\sf Win-Hyp}]
Let ${\sf G}=(\Ar,\WinObj_{\subeve},\WinObj_{\subadam})$ be a game, \eve\/ achieves $\WinObj_{\subeve}$ from state $\st$ under hypothesis $\WinObj_{\subadam}$ if there exists $\StrEve$ such that 
$$\outcome_{\st}(\StrEve) \subseteq \WinObj_{\subeve} \cup \overline{\WinObj_{\subadam}}.$$
\end{definition} 
\noindent
The synthesis rule in the definition above is called {\em winning under hypothesis}, {\sf Win-Hyp} for short.

\begin{example}
  \label{example:win-hyp-not-great}
  Let us consider the example of Fig.~\ref{fig:ex-adm} again. But now assume that the objective of \adam\/ is to visit $3$ infinitely often, i.e., $\WinObj_{\subadam}=\{ \pi \in \Plays \mid \pi \models \Box \Diamond 3 \}$. In this case, it should be clear then the strategy $1 \rightarrow 2$ and $3 \rightarrow 4$ for \eve\/ is winning for the objective 
$${\textsf{Win-Hyp}_{\Box \Diamond 4 \vee \overline{\Box \Diamond 3}}}=\{ \pi \in \Plays \mid \pi \models \Box \Diamond 4 \} \cup \overline{\{ \pi \in \Plays \mid \pi \models \Box \Diamond 3 \}}$$
i.e., under the hypothesis that the outcome satisfies the objective of \adam.

Unfortunately, there are strategies of \eve\/ which are winning for the rule {\sf Win-Hyp} but which are not desirable. As an example, consider the strategy that in $1$ chooses to go to $5$. In that case, the objective of \adam\/ is unmet and so this strategy of \eve\/ is winning for ${\textsf{Win-Hyp}_{\Box \Diamond 4 \vee \overline{\Box \Diamond 3}}}$, but clearly such a strategy is not interesting as it excludes the possibility to meet the objective of \eve.
\end{example}

\section{Assume Admissible Synthesis}

To define the notion of admissible strategy, we first need to define when a strategy $\sigma$ is dominated by a strategy $\sigma'$. We will define the notion for \eve, the definition for \adam\/ is symmetric. 

Let $\StrEve$ and $\StrEve'$ be two strategies of \eve\/ in the game arena $\Ar$. We say that $\StrEve'$ {\em dominates} $\StrEve$ if the following two conditions hold:
  \begin{enumerate}
  \item $  \displaystyle \forall \StrAdam \in \Sigma_{\subadam} \cdot \outcome_{\StInit}(\StrEve,\StrAdam) \in \WinObj_{\subeve} \rightarrow \outcome_{\StInit}(\StrEve',\StrAdam) \in \WinObj_{\subeve}$
  \medskip
  \item $\displaystyle \exists \StrAdam \in \Sigma_{\subadam} \cdot \outcome_{\StInit}(\StrEve,\StrAdam) \notin \WinObj_{\subeve} \land \outcome_{\StInit}(\StrEve',\StrAdam) \in \WinObj_{\subeve}$
  \end{enumerate}
\noindent
So a strategy $\StrEve$ is dominated by $\StrEve'$ if $\StrEve'$ does as well as $\StrEve$ against any strategy of \adam\ (condition~$1$), and there exists a strategy of \adam\/ against which $\StrEve'$ does better than $\StrEve$ (condition~$2$). 

\begin{definition}[Admissible Strategy]
A strategy is admissible if there does not exist a strategy that dominates it.
\end{definition}

Let ${\sf G}=(\Ar,\WinObj_{\subeve},\WinObj_{\subadam})$ be a game, the set of admissible strategies for \eve\/ is noted ${\sf Adm}_{\subeve}$, and the set of admissible strategies for \adam\/ is denoted ${\sf Adm}_{\subadam}$.

Clearly, a rational player should not play a dominated strategy as there always exists some strategy that behaves strictly better than the dominated strategy. So, a rational player only plays admissible strategies.

\begin{example}
Let us consider again the example of Fig.~\ref{fig:ex-adm} with $\WinObj_{\subeve}=\{ \pi \in \Plays \mid \pi \models \Box \Diamond 4 \}$ and $\WinObj_{\subadam}=\{ \pi \in \Plays \mid \pi \models \Box \Diamond 3 \}$. We claim that the strategy $\StrEve$ that plays $1 \rightarrow 5$ is not admissible in $\Ar$ from state $1$. This is because the strategy $\StrEve'$ that plays $1 \rightarrow 2$ and $4 \rightarrow 3$ dominates this strategy. Indeed, while $\StrEve$ is always losing for the objective of \eve, the strategy $\StrEve'$ wins for this objective whenever \adam\/ eventually plays $2 \rightarrow 3$.
\end{example}

\begin{definition}[{\sf AA}]
Let ${\sf G}=(\Ar,\WinObj_{\subeve},\WinObj_{\subadam})$ be a game, \eve\/ achieves $\WinObj_{\subeve}$ from $\st$ under the hypothesis that \adam\/ plays admissible strategies if
$$\exists \StrEve \in {\sf Adm}_{\subeve} \cdot \forall  \StrAdam \in {\sf Adm}_{\subadam} \cdot \outcome_{\st}(\StrEve,\StrAdam) \in \WinObj_{\subeve}.$$
\end{definition} 

\begin{example}
Let us consider again the example of Fig.~\ref{fig:ex-adm} with $\WinObj_{\subeve}=\{ \pi \in \Plays \mid \pi \models \Box \Diamond 4 \}$ and $\WinObj_{\subadam}=\{ \pi \in \Plays \mid \pi \models \Box \Diamond 3 \}$. We claim that the strategy $\StrEve$ of \eve\/ that plays $1 \rightarrow 2$ and $4 \rightarrow 3$ is admissible (see previous example) and winning against all the admissible strategies of \adam. This is a consequence of the fact that the strategy of \adam\/ that always plays $2 \rightarrow 2$, and which is the only counter strategy of \adam\/ against $\StrEve$, is {\em not} admissible. Indeed, this strategy falsifies $\WinObj_{\subadam}$ while a strategy that always chooses $2 \rightarrow 3$ enforces the objective of \adam.
\end{example}

\begin{theorem}[\cite{berwanger07,BRS14,BRS-concur15}]
For all games ${\sf G}=(\Ar,\WinObj_{\subeve},\WinObj_{\subadam})$, if $\WinObj_{\subeve}$ and $\WinObj_{\subadam}$ are omega-regular sets of plays, then ${\sf Adm}_{\subeve}$ and ${\sf Adm}_{\subadam}$ are both non empty sets. 

The problem of deciding if a game ${\sf G}=(\Ar,\WinObj_{\subeve},\WinObj_{\subadam})$, where $\WinObj_{\subeve}$ and $\WinObj_{\subadam}$ are omega-regular sets of plays expressed as Muller objectives, satisfies 
$$\exists \StrEve \in {\sf Adm}_{\subeve} \cdot \forall  \StrAdam \in {\sf Adm}_{\subadam} \cdot \outcome_{\st}(\StrEve,\StrAdam) \in \WinObj_{\subeve}$$
\noindent
is {\sc PSpace-complete}.
\end{theorem}

\paragraph{Additional Results.}
The assume-admissible setting we present here relies on procedures for \emph{iterative elimination of dominated strategies} for multiple players which was studied in~\cite{berwanger07} on games played on graphs.
In this context, dominated strategies are repeatedly eliminated for each player.
Thus, with respect to the new set of strategies of its opponent, new strategies may become dominated, and will therefore be eliminated, and so on until the process stabilizes.
In~\cite{BRS14}, we studied the algorithmic complexity of this problem and proved that for games
with Muller objectives, deciding whether all outcomes compatible with iteratively admissible strategy profiles
satisfy an omega-regular objective defined by a Muller condition is {\sc PSpace-complete} and in $\textsc{UP}\cap \textsc{coUP}$ for the special case of B\"uchi objectives.

The assume-admissible rule introduced 
in~\cite{BRS-concur15} is also defined for multiple players and corresponds, roughly, to the 
first iteration of the elimination procedure. 
We additionally prove
that if players have B\"uchi objectives, then the rule can be decided in polynomial-time. 
One advantage of the assume-admissible rule is
the \emph{rectangularity} of the solution set: the set of strategy profiles that witness the rule
can be written as a product of sets of strategies for each player. In particular, this means that
a strategy witnessing the rule can be chosen separately for each player. Thus, the rule is
\emph{robust} in the sense that the players do not need to agree on a strategy profile, but only
on the admissibility assumption on each other.
In addition, we show in~\cite{BRS-concur15} that the rule is amenable to abstraction techniques:
we show how state-space abstractions can be used to check a sufficient condition for assume-admissible, only doing computations on the abstract state space.

\paragraph{Related Works.}
The rule  ``winning under hypothesis'' (\textsf{Win-Hyp})  and its weaknesses 
are discussed in~\cite{BloemEJK14}. We have illustrated the limitations of this rule in Example~\ref{example:win-hyp-not-great}.

There are related works in the literature which propose concepts to model systems composed of several parts, each having their own objectives. The solutions that are proposed are based on $n$-players {\em non}-zero sum games. This is the case both for {\em assume-guarantee synthesis}~\cite{CH07} ({\sf AG}), and for {\em rational synthesis}~\cite{FismanKL10} ({\sf RS}). 

For the case of two player games, {\sf AG} is based on the concept of {\em secure equilibria}~\cite{KHJ06} ({\sf SE}), a refinement of Nash equilibria~\cite{nash50} ({\sf NE}).
In {\sf SE}, objectives of the players are lexicographic: each player first tries to force his own objective, and then tries to falsify the objectives of the other players. It was shown in~\cite{KHJ06} that {\sf SE} are the {\sf NE} that form enforceable contracts between the two players. 
When the {\sf AG} rule is extended to several players, as in~\cite{CH07}, it no longer corresponds to secure equilibria. We gave a direct algorithm for multiple players in~\cite{BRS-concur15}. The difference between {\sf AG} and {\sf SE} is that {\sf AG} strategies have to be resilient to deviations of all the other players, while {\sf SE} profiles have to be resilient to deviations by only one player.  A variant of the rule {\sf AG}, called {\em Doomsday equilibria}, has been proposed in~\cite{CDFR14}. We have also studied quantitative extensions of the notion of secure equilibria in~\cite{BruyereMR14}.

In the context of infinite games played on graphs, one well known limitation of {\sf NE} is the existence of {\em non-credible threats}. Refinements of the notion of {\sf NE}, like {\em sub-game perfect equilibria} ({\sf SPE}), have been proposed to overcome this limitation. {\sf SPE} for games played on graphs have been studied in e.g.,~\cite{Ummels06,BrihayeBMR15}. Admissibility does not suffer from this limitation. 

In {\sf RS}, the system is assumed to be monolithic and the environment is made of several components that are only {\em partially controllable}. In {\sf RS}, we search for a profile of strategies in which the system forces its objective and the players that model the environment are given an ``{\em acceptable}'' strategy profile, from which it is assumed that they will not deviate.
``Acceptable'' can be formalized by any {\em solution concept}, e.g., by {\sf NE}, 
\emph{dominant} strategies, or \emph{sub-game perfect equilibria}. This is the existential flavor of {\sf RS}. More recently, Kupferman et al. have proposed in~\cite{KupfermanPV14} a {\em universal} variant of this rule. In this variant, we search for a strategy of the system such that in all strategy profiles that extend this strategy for the system and that are {\sf NE}, the outcome of the game satisfies the specification of the system.

In~\cite{Faella09}, Faella studies several alternatives to the notion of winning strategy including the notion of admissible strategy. His work is for two-players but only the objective of one player is taken into account, the objective of the other player is left unspecified. In that work, the notion of admissibility is used to define a notion of {\em best-effort} in synthesis.

The notion of admissible strategy is definable in strategy logics~\cite{ChatterjeeHP10,MogaveroMV10} and decision problems related to the assume-admissible rule can be reduced to satisfiability queries in such logics. This reduction does not lead to worst-case optimal algorithms; we presented worst-case optimal algorithms in~\cite{BRS-concur15} based on our previous work~\cite{BRS14}.

\section{Regret Minimization}

In the previous section, we have shown how the notion of admissible strategy can
be used to relax the classical worst-case hypothesis made on the environment. In
this section, we review another way to relax this worst-case hypothesis. 

The idea is simple and intuitive. When looking for a strategy, instead of trying
to find a strategy which is worst-case optimal, we search for a strategy that
takes {\em best-responses} (against the behavior of the environment) as a yardstick.
That is, we would like to find a strategy that behaves ``not far'' from an optimal
response to the strategy of the environment\,---\,when the latter is fixed. The
notion of regret minimization is naturally defined in a quantitative setting
(although it also makes sense in a Boolean setting).

Let us now formally define the notion of regret associated to a
strategy of \eve. This definition is parameterized by a set of strategies for
\adam. 

\begin{definition}[Relative Regret]
Let $\Ar=\warena$ be a weighted game arena, let $\StrEve$ be a strategy of \eve,
the regret of this strategy relative to a set of strategies ${\sf
Str}_{\subadam} \subseteq \Sigma_{\subadam}$ is defined as follows:
$${\sf Reg}(\StrEve,{\sf Str}_{\subadam})=\sup_{\StrAdam \in {\sf
Str}_{\subadam}}\sup_{\StrEve' \in \Sigma_{\subeve}} \val(\StrAdam,\StrEve'
)-\val(\StrAdam,\StrEve).$$
\end{definition}

We interpret the sub-expression $\sup_{\StrEve' \in
\Sigma_{\subeve}} \val(\StrAdam,\StrEve')$ as the
{\em best-response} of \eve\/ against $\StrAdam$. Then, the relative regret
of a strategy of \eve\/ can be seen as the supremum of the differences between the
value achieved by $\StrEve$ against a strategy of \adam\/ and the value
achieved by the corresponding best-response.

We are now equipped to formally define the problem under study, which is
parameterized by payoff function ${\sf Val}(\cdot)$ and a set ${\sf
Str}_{\subadam}$ of strategies of \adam\/.

\begin{definition}[Regret Minimization]
Given a weighted game arena $\Ar$ and a rational threshold $r$, decide if there
exists a strategy $\StrEve$ for \eve\ such that
\[  {\sf Reg}(\StrEve,{\sf Str}_{\subadam}) \le r\]
and synthesize such a strategy if one exists.
\end{definition}

In~\cite{hpr15}, we have considered several types of strategies for \adam:
	the set $\Sigma_{\subadam}$, i.e., any strategy,
	the set $\Sigma_{\subadam}^{\sf ML}$, i.e., memoryless strategies for \adam, and
	the set $\Sigma_{\subadam}^{\sf W}$, i.e., word strategies for
	\adam.\footnote{To define word strategies, it is convenient to consider
		game arenas where edges have labels called letters.
		In that case, when playing a word strategy, \adam\/ commits to a
		sequence of letters (i.e., a word) and plays that word
		regardless of the exact state of the game. Word strategies are
		formally defined in~\cite{hpr15} and below.}
We will illustrate each of these cases on examples below.

\begin{example}
Let us consider the weighted game arena of Fig.~\ref{fig:ex-MP}, and let us
assume that we want to synthesize a strategy for \eve\/ that minimizes her
mean-payoff
regret against \adam\/ playing a memoryless strategy.  The memoryless
restriction is useful when designing a system that needs to perform well in an
environment which is only partially known. In practice, a controller may
discover the environment with which it is interacting during run-time. Such a
situation can be modeled by an arena in which choices in nodes of the
environment model an entire family of environments and each memoryless strategy
models a specific environment of the family. In such cases, if we want to design
a controller that performs reasonably well against all the possible
environments, we can consider each best-response of \eve\/ for each environment
and then try to choose one unique strategy for \eve\/ that minimizes the
difference in performance w.r.t. those best-responses: a regret-minimizing
strategy.

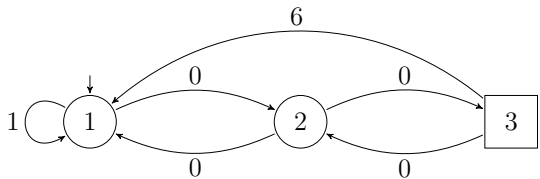
\begin{figure}[thb]
 \begin{center}
  
\scalebox{0.7}{\begin{tikzpicture}[->,>=stealth',shorten >=1pt,shorten <=1pt,auto,node
    distance=2.5cm,bend angle=45,font=\Large]
    \tikzstyle{p1}=[draw,circle,text centered,minimum size=10mm]
    \tikzstyle{p2}=[draw,rectangle,text centered,minimum size=10mm]
    \tikzstyle{empty}=[]
    \node[p1,initial,initial where=above,initial text=] (1) at (0,0) {$1$};
    \node[p1] (2) at (4,0) {$2$};
    \node[p2] (3) at (8,0) {$3$};

\path[->] (1) edge [loop left, out=150, in=210,looseness=2, distance=12mm] node [left] {$1$} (1);
	\path[->] (3) edge[bend right=40,swap] node {$6$} (1);
	\path[->] (1) edge[bend left=25] node {$0$} (2);
	\path[->] (2) edge[bend left=25] node {$0$} (3);
	\path[->] (2) edge[bend left=25] node {$0$} (1);
	\path[->] (3) edge[bend left=25] node {$0$} (2);
      \end{tikzpicture}}
  \end{center}
  \vspace{-4mm}
\caption{An example of a two-player game arena with {\sf MP} objective for \eve. Rounded positions belong to \eve, and squared positions belong to \adam.\label{fig:ex-MP}}
\end{figure}

In our example, prior to a first visit to state $3$, we do not know if the edge
$3 \rightarrow 2$ or the edge $3 \rightarrow 1$ will be activated by \adam. But
as \adam\/ is bound to play a memoryless strategy, once he has chosen one of the
two edges, we know that he will stick to this choice. 

A regret-minimizing strategy in this example is as follows: play $1
\rightarrow 2$, then  $2 \rightarrow 3$, if \adam\/ plays $3 \rightarrow 2$,
then play $2 \rightarrow 1$ and then $1 \rightarrow 1$ forever, otherwise
\adam\/ plays $3 \rightarrow 1$ and then \eve\/ should continue to play $1
\rightarrow 2$ and  $2 \rightarrow 3$ forever. This strategy has regret $0$.
Note that this strategy uses memory and that there is no memoryless strategy of
\eve\/ with regret $0$ in this game.
\end{example}

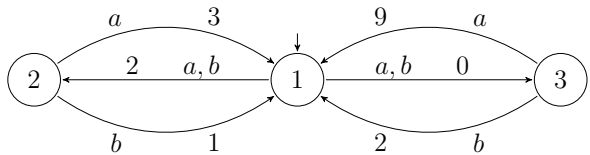
\begin{figure}
 \begin{center}
  \scalebox{0.7}{
\begin{tikzpicture}[->,>=stealth',shorten >=1pt,shorten <=1pt,auto,node
    distance=3cm,bend angle=45,font=\Large]
    \tikzstyle{p1}=[draw,circle,text centered,minimum size=10mm]
    \tikzstyle{p2}=[draw,rectangle,text centered,minimum size=10mm]

\node[p1,initial,initial where=above,initial text=] (q1) at (0,0) {$1$};
\node[p1] (q2) at (-5,0) {$2$};
\node[p1] (q3) at (5,0) {$3$};

\path[->] (q1) edge[swap] node[pos=0.33,inner ysep=0pt] {$a,b$} node[pos=0.66]{$2$} (q2);
\path[->] (q1) edge node[pos=0.33,inner ysep=0pt] {$a,b$} node[pos=0.66] {$0$} (q3);
\path[->] (q2) edge[bend right=35,swap] node[pos=0.33] {$b$} node[pos=0.66] {$1$} (q1);
\path[->] (q3) edge[bend left=35] node[pos=0.33] {$b$} node[pos=0.66] {$2$} (q1);
\path[->] (q2) edge[bend left=35] node[pos=0.33] {$a$} node[pos=0.66] {$3$} (q1);
\path[->] (q3) edge[bend right=35,swap] node[pos=0.33] {$a$} node[pos=0.66] {$9$} (q1);
\end{tikzpicture}
}
  \end{center}
  \vspace{-4mm}
\caption{An example of a two-player game arena with {\sf MP} objective for \eve. Edges are annotated by letters: \adam\/ chooses a word $w$ and \eve\/ resolves the non-determinism on edges.}\label{fig:regret_word}
\end{figure}

Let us now illustrate the interest of the notion of regret minimization when
\adam\/ plays {\em word} strategies. When considering this restriction, it is
convenient to consider letters that label the edges of the graph (Fig.~\ref{fig:regret_word}). A word
strategy for \adam\/ is a function $w : \mathbb{N} \rightarrow \{ a,b \}$. In
this setting \adam\/ plays a sequence of letters and this sequence is
independent of the current state of the game. We have shown in~\cite{hpr15} that the
notion of regret minimization relative to word strategies is a generalization of
the notion of {\em good-for-games automata} introduced by Henzinger and Piterman
in~\cite{hp06}.

\vspace{-4mm}
\vspace{-4mm}

\begin{example}
In this example, a strategy of \eve\/ determines how  to resolve non-determinism
in state $1$. The best strategy of \eve\/ for mean-payoff regret minimization is
to always take the edge $1 \rightarrow 3$. Indeed, let us consider all the
sequences of two letters that \adam\/ can choose and compute the regret of
choosing $1 \rightarrow 2$ ({\sf left}) and the regret of choosing $1
\rightarrow 3$ ({\sf right}):
   \begin{itemize}
   	\item $*a$ with $* \in \{a,b\}$, the regret of {\sf left} is equal to $0$,
		and the regret of {\sf right} is $\frac{5-3}{2}=1$.
	\item $*b$ with $* \in \{a,b\}$, the regret of {\sf left} is equal to
		$\frac{9-3}{2}=3$, and the regret of {\sf right} is $0$. 
   \end{itemize}
So the strategy that minimizes the regret of \eve\/ is to always take the arrow
$1 \rightarrow 3$ ({\sf right}), the regret is then equal to $1$.
\end{example}
 
In~\cite{hpr15}, we have studied the complexity of deciding the existence of
strategies for \eve\/ that have less than a given regret threshold.  The results
that we have obtained are summarized in the theorem below.

\begin{theorem}[\cite{hpr15}]
Let $\Ar=\warena$ be a weighted game arena, the complexity of deciding if \eve\/
has a strategy with regret less than or equal to a threshold $r \in \rat$
against \adam\/ playing:
  \begin{itemize}
	\item a strategy in $\Sigma_{\subadam}$, is {\sc PTime-Complete} for
		payoff functions ${\sf inf}$, ${\sf sup}$, ${\sf lim inf}$, ${\sf lim
		sup}$, and in {\sc NP}$\cap${\sc coNP} for {\sf MP}.
	\item a strategy in $\Sigma_{\subadam}^{\sf ML}$, is in {\sc PSpace} for
		payoff functions ${\sf inf}$, ${\sf sup}$, ${\sf lim inf}$, ${\sf lim
		sup}$, and ${\sf MP}$, and is {\sc  coNP-Hard} for ${\sf inf}$,
		${\sf sup}$, ${\sf lim sup}$, and {\sc PSpace-Hard} for ${\sf
		lim inf}$, and ${\sf MP}$.
	\item a strategy in $\Sigma_{\subadam}^{\sf W}$, is {\sc
		ExpTime-Complete} for payoff functions ${\sf inf}$, ${\sf sup}$, ${\sf
		lim inf}$, ${\sf lim sup}$, and undecidable for ${\sf MP}$.
  \end{itemize}
\end{theorem}
The above results are obtained by reducing the synthesis of regret-minimizing
strategies to finding winning strategies in classical games. For instance, a
strategy for \eve\/ that minimizes regret against $\Sigma_{\subadam}^{\sf ML}$
for the mean-payoff measure corresponds to finding a winning strategy in a
mean-payoff game played on a larger game arena which encodes the witnessed
choices of \adam\/ and forces him to play positionally. When minimizing regret
against word strategies, for the decidable cases the reduction is done to
parity games and is based on the \emph{quantitative simulation} games defined
in~\cite{CDH10}.

\paragraph*{Additional Results.}
Since synthesis of regret-minimizing strategies against word strategies of
\adam\/ is undecidable with measure {\sf MP}, we have considered the sub-case
which limits the amount of memory the desired controller can use (as
in~\cite{akl10}). That is, we ask whether there exists a strategy of \eve\/
which uses at most memory $m$ and ensures regret at most $r$. In~\cite{hpr15} we
showed that this problem is in {\sc NTime}$(m^2|\calA|^2)$ for {\sf MP}.

\begin{theorem}[\cite{hpr15}]
Let $\Ar=\warena$ be a weighted game arena, the complexity of deciding if \eve\/
has a strategy using memory of at most $m$ with regret less than or equal to a
threshold $\lambda \in \rat$ against \adam\/ playing a strategy in
$\Sigma_{\subadam}^{\sf W}$, is in non-deterministic polynomial time w.r.t. $m$
and $|\Ar|$ for ${\sf inf}$, ${\sf sup}$, ${\sf lim inf}$, ${\sf lim sup}$,
and {\sf MP}.
\end{theorem}

Finally, we have established the equivalence of a quantitative extension of the
notion of good-for-games automata~\cite{hp06} with
determinization-by-pruning of the refinement of an automaton~\cite{akl10} and
our regret games against word strategies of \adam. Before we can formally state
these results, some definitions are needed.

\begin{definition}[Weighted Automata]
	A finite weighted automaton is a tuple $\langle Q,q_{\sf
	init},A,\Delta,{\sf w}\rangle$ where: $Q$ is a finite set of states,
	$q_{\sf init} \in Q$ is the initial state, $A$ is a finite alphabet of
	actions or symbols, $\Delta \subseteq Q \times A \times Q$ is the
	transition relation, and ${\sf w} : \Delta \to \mathbb{Z}$ is the weight
	function.
\end{definition}

A \emph{run} of an automaton on a word $a \in A^\omega$ is an infinite sequence
of transitions $\rho = (q_0,a_0,q_1)(q_1,a_1,q_2)\dots \in \Delta^\omega$ such
that $q_0 = q_{\sf init}$ and $a_i = a(i)$ for all $i \ge 0$. As with plays in a
game, each run is assigned a \emph{value} with a payoff function ${\sf
Val(\cdot)}$. A weighted automaton $\mathcal{M}$ defines a function $A^\omega
\to \mathbb{R}$ by assigning to $a \in A^\omega$ the supremum over all the
values of its runs on $a$. The automaton is said to be \emph{deterministic} if
for all $q \in Q$ and $x \in A^\omega$ the set $\{ q' \in Q \mid (q,x,q') \in
\Delta \}$ is a singleton.

In~\cite{hp06}, Henzinger and Piterman introduced the notion of
\emph{good-for-games automata}. A non-deterministic automaton is good for
solving games if it fairly simulates the equivalent deterministic automaton.

\begin{definition}[$\alpha$-good-for-games]
	A finite weighted automaton $\mathcal{M}$ is 
	\emph{$\alpha$-good-for-games} if
	a player (Simulator), against any word $x \in A^\omega$ spelled by
	Spoiler, can resolve non-determinism in $\mathcal{M}$ so that
	the resulting run has value $v$ and $\mathcal{M}(x) - v \le \alpha$.
\end{definition}
The above definition is a quantitative generalization of the notion proposed
in~\cite{hp06}. We link their class of automata with our regret games in the
sequel.
\begin{proposition}[\cite{hpr15}]\label{pro:rel-gfg}
	A weighted automaton $\mathcal{M} = \langle Q,q_{\sf init},A,\Delta,{\sf
	w}\rangle$ is $\alpha$-good-for-games if and only if there exists a
	strategy $\sigma_{\subeve}$ for \eve\/ with relative regret of at most
	$\alpha$ against strategies
	$\Sigma_{\subadam}^{\sf W}$ of \adam\/.
\end{proposition}

Our definitions also suggest a natural notion of approximate determinization
for weighted automata on infinite words. This is related to recent work by
Aminof et al.: in~\cite{akl10}, they introduce the notion of {\em
approximate-determinization-by-pruning} for weighted sum automata over finite
words. For $\alpha \in (0,1]$, a weighted sum automaton is {\em
$\alpha$-determinizable-by-pruning} if there exists a finite state strategy to
resolve non-determinism and that constructs a run whose value is at least
$\alpha$ times the value of the maximal run of the given word. So, they consider
a notion of approximation which is a {\em ratio}. Let us introduce some
additional definitions required to formalize the notion of
determinizable-by-pruning.

Consider two weighted automata $\mathcal{M} = \langle Q,q_{\sf
init},A,\Delta,{\sf w}\rangle$ and $\mathcal{M'} = \linebreak \langle Q',q'_{\sf
init},A,\Delta',{\sf w'}\rangle$. We say that \emph{$\mathcal{M'}$
$\alpha$-approximates $\mathcal{M}$} if $|\mathcal{M}(x) - \mathcal{M'}(x)| \le
\alpha$, for all $x \in A^\omega$. We say that $\mathcal{M}$ \emph{embodies}
$\mathcal{M'}$ if $Q' \subseteq Q$, $\Delta' \subseteq \Delta$, and ${\sf w'}$
agrees with ${\sf w}$ on $\Delta'$. For an integer $k \ge 0$, the $k$-refinement
of $\mathcal{M}$ is the automaton obtained by refining the state-space of
$\mathcal{M}$ using $k$ boolean variables.

\begin{definition}[$(\alpha,k)$-determinizable-by-pruning]
	A finite weighted au\-to\-ma\-ton $\mathcal{M}$ is
	\emph{$(\alpha,k)$-determinizable-by-pruning} if the $k$-refinement of
	$\mathcal{M}$ embodies a deterministic automaton which
	$\alpha$-approximates $\mathcal{M}$.
\end{definition}

We show in~\cite{hpr15} that when \adam\/ plays word strategies only, our notion
of regret defines a notion of approximation with respect to the {\em difference}
metric for weighted automata (as defined above).

\begin{proposition}[\cite{hpr15}]\label{pro:rel-dbp}
	A weighted automaton $\mathcal{M} = \langle Q,q_{\sf init},A,\Delta,{\sf
	w}\rangle$ is $\alpha$-determinizable-by-pruning if and only if there exists a
	strategy $\sigma_{\subeve}$ for \eve\/ using memory at most $2^m$ with
	relative regret of at most
	$\alpha$ against strategies $\Sigma_{\subadam}^{\sf W}$ of \adam\/.
\end{proposition}

\paragraph{Related Works}
The notion of regret minimization is important in game and decision theory, see
e.g.,~\cite{zjbp08} and additional bibliographical pointers there.  The concept of {\it iterated} regret
minimization has been recently proposed by Halpern et al. for {\em
non-zero} sum games~\cite{hp12}. In~\cite{fgr10}, the concept is applied to games played on weighted graphs with
shortest path objectives. Variants on the different sets of strategies considered for \adam\/ were not considered there.

In~\cite{df11}, Damm and Finkbeiner introduce the notion of \emph{remorse-free strategies}. The notion is introduced in order to define a notion of {\em best-effort} strategy when winning strategies do not exist. Remorse-free strategies are exactly the strategies which minimize regret in games with $\omega$-regular
objectives in which the environment (\adam) is playing word strategies only. 
The authors of~\cite{df11} do not establish lower bounds on the complexity of
the realizability and synthesis problems for remorse-free strategies.

A concept equivalent to good-for-games automata is that of
history-\linebreak determinism~\cite{colcombet12}. Proposition~\ref{pro:rel-gfg} thus
allows us to generalize history-determinism to a quantitative setting via this
relationship with good-for-games automata.

Finally, we would like to highlight some differences between our work and the study of
Aminof et al. in~\cite{akl10} on determinization-by-pruning. First, we consider
infinite words while they consider finite words. Second, we study a general
notion of regret minimization problem in which \eve\/ can use any strategy while
they restrict their study to fixed memory strategies only and leave the problem
open when the memory is not fixed a priori.

\section{Game Arenas with Expected Adversary}

In the two previous sections we have relaxed the worst-case hypothesis on the environment (modeled by the behavior of \adam) by either considering an explicit objective for the environment or by considering  as yardsticks the best-responses to the strategies of \adam. Here, we introduce another model where the environment is modeled as a stochastic process (i.e., \adam\/ is expected to play according to some known \textit{randomized} strategy) and we are looking for strategies for \eve\/ that ensure good expectation against this stochastic process while guaranteeing acceptable worst-case performance even if \adam\/ deviates from his expected behavior.

To define formally this new framework, we need game arenas in which an expected behavior for \adam\/ is given as a \textit{memoryless randomized strategy}.\footnote{It should be noted that we can easily consider finite-memory randomized strategies for \adam, instead of memoryless randomized strategies. This is because we can always take the synchronized product of a finite-memory randomized strategy with the game arena to obtain a new game arena in which the finite-memory strategy on the original game arena is now equivalent to a memoryless strategy.} We first introduce some notation. Given a set $A$, let~$\calD(A)$ denote the set of rational probability distributions over~$A$,
and, for $d \in \calD(A)$, we denote its support by~$\supp(d) = \{a \in A \mid d(a) > 0 \} \subseteq A$.
\begin{definition}
Fix a weighted game arena $\Ar=\warena$. A memoryless randomized strategy for \adam\/ is a function 
$$\StrAdam^{{\sf rnd}} : \StAdam \rightarrow \calD( \states )$$
\noindent
such that for all $\st \in \StAdam$, $\supp(\StrAdam^{{\sf rnd}}(\st)) \subseteq \{ \st' \in \states \mid (\st, \st') \in \Edges \}$.
\end{definition}

For the rest of this section, we model the expected behavior of \adam\/ with a strategy $\StrAdam^{{\sf rnd}}$, given as part of the input for the problem we will consider. Given a weighted game arena $\Ar$ and a memoryless randomized strategy $\StrAdam^{{\sf rnd}}$ for \adam\/, we are left with a model with both non-deterministic choices (for \eve\/) and stochastic transitions (due to the randomized strategy of \adam\/). This is essentially what is known in the literature as a $1\frac{1}{2}$-player game or more commonly, a \textit{Markov Decision Process} (MDP), see for example~\cite{Puterman94,filar1997}. One can talk about plays, strategies and other notions in MDPs as introduced for games.

\begin{figure}[t]
  \centering
  \scalebox{0.7}{\begin{tikzpicture}[->,>=stealth',shorten >=1pt,shorten <=1pt,auto,node
    distance=2.5cm,bend angle=45,font=\Large]
    \tikzstyle{p1}=[draw,circle,text centered,minimum size=10mm]
    \tikzstyle{p2}=[draw,rectangle,text centered,minimum size=10mm]
    \tikzstyle{empty}=[]
    \node[p1,initial,initial where=above,initial text=] (1) at (0,0) {$1$};
    \node[p1] (2) at (4,0) {$2$};
    \node[p2] (3) at (8,0) {$3$};

\path[->] (1) edge [loop left, out=150, in=210,looseness=2, distance=12mm] node [left] {$1$} (1);
	\path[->] (3) edge[bend right=50] node {$6$} node[swap,pos=0.025] {$\frac{9}{10}$} (1);
	\path[->] (1) edge[bend left=25] node {$0$} (2);
	\path[->] (2) edge[bend left=25] node {$0$} (3);
	\path[->] (2) edge[bend left=25] node {$0$} (1);
	\path[->] (3) edge[bend left=25] node {$0$} node[pos=0.15] {$\frac{1}{10}$} (2);
      \end{tikzpicture}}
  \vspace{-2mm}
      \caption{A game arena associated with a memoryless randomized strategy for \adam\/ can be seen as an MDP: the fractions represent the respective probability to take each outgoing edge when leaving state $3$.}\label{fig:ex-MP-exp}
\end{figure}
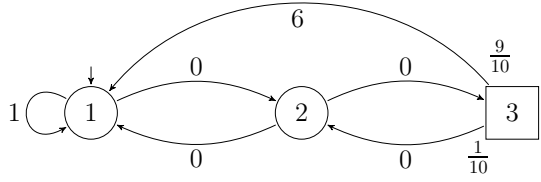

Consider the game in Fig.~\ref{fig:ex-MP-exp}. We can see it as a classical two-player game if we forget about the fractions around state $3$. Now assume that we fix the memoryless randomized strategy $\StrAdam^{{\sf rnd}}$ for \adam\/ to be the one that, from~$3$, goes to $1$ with probability $\frac{9}{10}$ and to $2$ with the remaining probability, $\frac{1}{10}$. This is represented by the fractions on the corresponding outgoing edges. In the remaining model, only \eve\/ still has to pick a strategy: it is an MDP. We denote this MDP by $\Ar[\StrAdam^{{\sf rnd}}]$.

Let us go one step further. Assume now that \eve\/ also picks a strategy $\StrEve$ in this MDP. Now we obtain a fully stochastic process called a \textit{Markov Chain} (MC). We denote it by $\Ar[\StrEve,\StrAdam^{{\sf rnd}}]$. In an MC, an \textit{event} is a measurable set of plays. It is well-known from the literature~\cite{Vardi-focs85} that every event has a uniquely defined probability (Carath\'eodory's extension theorem induces a unique probability measure on the Borel $\sigma$-algebra over plays in the MC). Given $\event$ a set of plays in $M = \Ar[\StrEve,\StrAdam^{{\sf rnd}}]$, we denote by $\mathbb{P}_{M}(\event)$ the probability that a play belongs to $\event$ when $M$ is executed for an infinite number of steps. Given a measurable value function $\val$, we denote by $\expect_{M}(\val)$ the \textit{expected value} or \textit{expectation} of $\val$ over plays in $M$. In this paper, we focus on the mean-payoff function $\MP$.

We are now finally equipped to formally define the problem under study.

\begin{definition}[Beyond Worst-Case Synthesis]
Given a weighted game arena $\Ar$, a stochastic model of \adam\/ given as a memoryless randomized strategy $\StrAdam^{{\sf rnd}}$, and two rational thresholds $\lambda_{{\sf wc}},\lambda_{{\sf exp}}$, decide if there exists a strategy $\StrEve$ for \eve\/ such that
$$\begin{cases}
     \forall \pi \in \outcome^{\Ar}_{\initState}(\StrEve) \cdot \val(\pi) > \lambda_{\sf wc} \\
     \expect_{\Ar[\StrEve,\StrAdam^{{\sf rnd}}]}(\val) > \lambda_{\sf exp} 
    \end{cases}$$
and synthesize such a strategy if one exists.
\end{definition}

Intuitively, we are looking for strategies that can \textit{simultaneously} guarantee a worst-case performance higher than $\lambda_{{\sf wc}}$, i.e., against any behavior of \adam\/ in the game $\Ar$, \textit{and} guarantee an expectation higher than $\lambda_{{\sf exp}}$ when faced to the expected behavior of \adam\/, i.e., when played in the MDP $\Ar[\StrAdam^{{\sf rnd}}]$.
We can of course assume w.l.o.g. that $\lambda_{\sf wc} < \lambda_{\sf exp}$,
otherwise the problem reduces trivially to just a worst-case requirement: any lower bound on the worst-case value is also a lower bound on the expected value.

\begin{example}
Consider the arena depicted in Fig.~\ref{fig:ex-MP-exp}. As mentioned before, the probability distribution models the expected behavior of \adam. Assume that we want now to synthesize a strategy for \eve\/ which ensures that $(C_1)$ the mean-payoff will be at least $\frac{1}{3}$ no matter how \adam\/ behaves (worst-case guarantee), and $(C_2)$ at least $\frac{3}{2}$ if \adam\/ plays according to his expected behavior (good expectation).

First, let us study whether this can be achieved through the two classical solution concepts used in games and MDPs respectively. We start by considering the arena as a traditional two-player zero-sum game: in this case, it is known that an optimal memoryless strategy exists~\cite{EM79}. Let $\StrEve^{\sf wc}$ be the strategy of \eve\/ that always plays $1 \rightarrow 1$ and $2 \rightarrow 1$. That strategy maximizes the worst-case mean-payoff, as it enforces a mean-payoff of $1$ no matter how \adam\/ behaves. Thus, $(C_1)$ is satisfied. Observe that if we consider the arena as an MDP (i.e., taking the probabilities into account), this strategy yields an expected value of $1$ as the unique possible play from state $1$ is to take the self-loop forever. Hence this strategy does not satisfy $(C_2)$.

Now, consider the arena as an MDP. Again, it is known that the expected value can be maximized by a memoryless strategy~\cite{Puterman94,filar1997}. Let $\StrEve^{\sf exp}$ be the strategy of \eve\/ that always chooses the following edges: $1 \rightarrow 2$ and $2 \rightarrow 3$. Its expected mean-payoff can be calculated in two steps: first computing the probability vector that represents the limiting stationary distribution of the irreducible MC induced by this strategy, second multiplying it by the vector containing the expected weights over outgoing edges for each state. In this case, it can be shown that the expected value is equal to $\frac{54}{29}$, hence the strategy does satisfy $(C_2)$. Unfortunately, it is clearly not acceptable for $(C_1)$ as, if \adam\/ does not behave according to the stochastic model and always chooses to play $3 \rightarrow 2$, the mean-payoff will be equal to zero.

Hence this shows that the classical solution concepts do not suffice if one wants to go beyond the worst-case and mix guarantees on the worst-case and the expected performance of strategies. In contrast, with the framework developed in~\cite{bruyere_STACS2014,DBLP:journals/corr/BruyereFRR14}, it is indeed possible for the considered arena (Fig.~\ref{fig:ex-MP-exp}) to build a strategy for \eve\/ that ensures the worst-case constraint $(C_1)$ and at the same time, yields an expected value arbitrarily close to the optimal expectation achieved by strategy~$\StrEve^{\sf exp}$. In particular, one can build a finite-memory strategy that guarantees \textit{both} $(C_1)$ and $(C_2)$. The general form of such strategies is a combination of $\StrEve^{\sf exp}$  and $\StrEve^{\sf wc}$ in a well-chosen pattern. Let $\StrEve^{\sf cmb(K, L)}$ be a \textit{combined strategy} parameterized by two integers $K, L \in \nat$. The strategy is as follows.
  \begin{enumerate}
  	\item Play according to $\StrEve^{\sf exp}$ for $K$ steps.
	\item If the mean-payoff over the last $K$ steps is larger than the worst-case threshold $\lambda_{\sf wc}$ (here $\frac{1}{3}$),
	then go to phase 1.
	\item Otherwise, play according to $\StrEve^{\sf wc}$ for $L$ steps, and then go to phase 1.
  \end{enumerate}

Intuitively, the strategy starts by mimicking $\StrEve^{\sf exp}$ for a long time,
and the witnessed mean-payoff over the $K$ steps will be close to the optimal expectation with high probability.
Thus, with high probability it will be higher than $\lambda_{\sf exp}$, and therefore higher than $\lambda_{\sf wc}$\,---\,recall that we assumed $\lambda_{\sf wc} < \lambda_{\sf exp}$.
If this is not the case, then \eve\/ has to switch to $\StrEve^{\sf wc}$ for sufficiently many steps $L$ in order to make sure that the worst-case constraint $(C_1)$ is satisfied before switching back to $\StrEve^{\sf exp}$.

One of the key results of~\cite{bruyere_STACS2014} is to show that for any $\lambda_{\sf wc} < \mu$, where $\mu$ denotes the optimal worst-case value guaranteed by $\StrEve^{\sf wc}$, and for any expected value threshold $\lambda_{{\sf exp}} < \nu$, where $\nu$ denotes the optimal expected value guaranteed by $\StrEve^{\sf exp}$, it is possible to compute values for $K$ and $L$ such that $\StrEve^{\sf cmb(K, L)}$ satisfies the beyond worst-case constraint for thresholds $\lambda_{\sf wc}$ and $\lambda_{{\sf exp}}$. For instance, in the example, where $\lambda_{\sf wc} = \frac{1}{3} < 1$ and $\lambda_{\sf exp} = \frac{3}{2} < \frac{54}{29}$, one can compute appropriate values of the parameters following the technique presented in~\cite[Theorem 5]{bruyere_STACS2014}. The crux is proving that, for large enough values of $K$ and $L$, the contribution to the expectation of the phases when $\StrEve^{\sf cmb(K, L)}$ mimics $\StrEve^{\sf wc}$ are negligible,
and thus the expected value yield by $\StrEve^{\sf cmb(K, L)}$ tends to the optimal one given by $\StrEve^{\sf exp}$, while at the same time the strategy ensures that the worst-case constraint is met.
\end{example}

In the next theorem, we sum up some of the main results that we have obtained for the beyond worst-case synthesis problem applied to the mean-payoff value function.

\begin{theorem}[\cite{bruyere_STACS2014,DBLP:journals/corr/BruyereFRR14,DBLP:conf/lics/ClementeR15}]
The beyond worst-case synthesis problem for the mean-payoff is in {\sc NP}~$\cap$~{\sc coNP}, and at least as hard as deciding the winner in two-player zero-sum mean-payoff games, both when looking for finite-memory or infinite-memory strategies of \eve. When restricted to finite-memory strategies, pseudo-polynomial memory is both sufficient and necessary.
\end{theorem}

The {\sc NP}~$\cap$~{\sc coNP}-membership is good news as it matches the long-standing complexity barrier for two-player zero-sum mean-payoff games~\cite{EM79,ZP96,BCDGR11,DBLP:journals/iandc/Chatterjee0RR15}: the beyond worst-case framework offers \textit{additional modeling power for free} in terms of decision complexity. It is also interesting to note that in general, infinite-memory strategies are more powerful than finite-memory ones in the beyond worst-case setting, which is not the case for the classical problems in games and MDPs.

Looking carefully at the techniques from~\cite{bruyere_STACS2014,DBLP:journals/corr/BruyereFRR14}, it can be seen that the main bottleneck in complexity is solving mean-payoff games in order to check whether the worst-case constraint can be met. Therefore, a natural relaxation of the problem is to consider the \textit{beyond almost-sure threshold problem} where the worst-case constraint is softened by only asking that a threshold is satisfied with probability one against the stochastic model given as the strategy $\StrAdam^{{\sf rnd}}$ of \adam. In this case, the complexity is reduced.

\begin{theorem}[\cite{DBLP:conf/lics/ClementeR15}]
The beyond almost-sure threshold problem for the mean-payoff is in {\sc PTime} and finite-memory strategies are sufficient.
\end{theorem}

\paragraph{Related Works} We originally introduced the beyond worst-case framework in~\cite{bruyere_STACS2014} where we studied both mean-payoff and shortest path objectives. This framework generalizes classical problems for two-player zero-sum games and MDPs. In mean-payoff games, optimal memoryless strategies exist and deciding the winner lies in {\sc NP}~$\cap$~{\sc coNP} while no polynomial algorithm is known~\cite{EM79,ZP96,BCDGR11,DBLP:journals/iandc/Chatterjee0RR15}. For shortest path games, where we consider game graphs with strictly positive weights and try to minimize the accumulated cost to target, it can be shown that memoryless strategies also suffice, and the problem is in {\sc PTime}~\cite{DBLP:dblp_journals/mst/KhachiyanBBEGRZ08}. In MDPs, optimal strategies for the expectation are studied in~\cite{Puterman94,filar1997} for the mean-payoff and the shortest path: in both cases, memoryless strategies suffice and they can be computed in {\sc PTime}. While we saw that the beyond worst-case synthesis problem does not cost more than solving games for the mean-payoff, it is not the case anymore for the shortest path: we jump from {\sc PTime} to a pseudo-polynomial-time algorithm. We proved in~\cite[Theorem 11]{bruyere_STACS2014} that the problem is inherently harder as it is {\sc NP}-hard.

The beyond worst-case framework was extended to the multi-dimensional setting\,---\,where edges are fitted with vectors of integer weights\,---\,in~\cite{DBLP:conf/lics/ClementeR15}. The general case is proved to be {\sc coNP}-complete.

Our strategies can be considered as {\em strongly risk averse}: they avoid at all cost outcomes that are below a given threshold (no matter what is their probability), and inside the set of those {\em safe} strategies, we maximize the expectation. Other different notions of risk have been studied for MDPs: in~\cite{WL99}, the authors want to find policies which minimize the probability (risk) that the total discounted rewards do not exceed a specified value (target); in~\cite{FKR95} the authors want policies that achieve a specified value of the long-run limiting average reward at a specified probability level (percentile). The latter problem was recently extended significantly in the framework of \textit{percentile queries}, which provide elaborate guarantees on the performance profile of strategies in multi-dimensional MDPs~\cite{RRS-cav15}. While all those strategies limit risk, they only ensure {\em low probability} for bad behaviors but they do not ensure their absence, furthermore, they do not ensure good expectation either.

Another body of work is the study of strategies in MDPs that achieve a trade-off between the expectation and the variance over the outcomes (e.g.,~\cite{brazdil_LICS2013} for the mean-payoff,~\cite{mannor_ICML2011} for the cumulative reward), giving a statistical measure of the stability of the performance. In our setting, we strengthen this requirement by asking for \textit{strict guarantees on individual outcomes}, while maintaining an appropriate expected payoff.

A survey of rich behavioral models extending the classical approaches for MDPs\,---\,including the beyond worst-case framework presented here\,---\,was published in~\cite{DBLP:conf/vmcai/RandourRS15}, with a focus on the shortest path problem.

\bibliographystyle{abbrv}
\bibliography{refs}

\begin{thebibliography}{10}

\bibitem{akl10}
B.~Aminof, O.~Kupferman, and R.~Lampert.
\newblock Reasoning about online algorithms with weighted automata.
\newblock {\em ACM Transactions on Algorithms}, 2010.

\bibitem{BK08}
C.~Baier and J.-P. Katoen.
\newblock {\em Principles of model checking}.
\newblock {MIT} Press, 2008.

\bibitem{berwanger07}
D.~Berwanger.
\newblock Admissibility in infinite games.
\newblock In {\em Proc. of {STACS}}, LNCS 4393, pages 188--199. Springer, 2007.

\bibitem{BloemEJK14}
R.~Bloem, R.~Ehlers, S.~Jacobs, and R.~K{\"{o}}nighofer.
\newblock How to handle assumptions in synthesis.
\newblock In {\em Proc. of {SYNT}}, {EPTCS} 157, pages 34--50, 2014.

\bibitem{adam2008admissibility}
A.~Brandenburger, A.~Friedenberg, and H.~J. Keisler.
\newblock Admissibility in games.
\newblock {\em Econometrica}, 76(2), 2008.

\bibitem{brazdil_LICS2013}
T.~Br{\'a}zdil, K.~Chatterjee, V.~Forejt, and A.~Kucera.
\newblock Trading performance for stability in {M}arkov decision processes.
\newblock In {\em Proc. of LICS}, pages 331--340. IEEE, 2013.

\bibitem{lata16}
R.~Brenguier, L.~Clemente, P.~Hunter, G.~A. P\'{e}rez, M.~Randour, J.-F.
  Raskin, O.~Sankur, and M.~Sassolas.
\newblock Non-zero sum games for reactive synthesis.
\newblock In {\em Proc. of LATA}, LNCS. Springer, 2016.
\newblock To appear.

\bibitem{BRS-concur15}
R.~Brenguier, J.-F. Raskin, and O.~Sankur.
\newblock {Assume-admissible synthesis}.
\newblock In {\em Proc. of CONCUR}, LIPIcs 42, pages 100--113. Schloss
  Dagstuhl--LZI, 2015.

\bibitem{BRS14}
R.~Brenguier, J.-F. Raskin, and M.~Sassolas.
\newblock The complexity of admissibility in omega-regular games.
\newblock In {\em Proc. of {CSL-LICS}}, pages 23:1--23:10. ACM, 2014.

\bibitem{BrihayeBMR15}
T.~Brihaye, V.~Bruy{\`{e}}re, N.~Meunier, and J.-F. Raskin.
\newblock Weak subgame perfect equilibria and their application to quantitative
  reachability.
\newblock In {\em Proc. of CSL}, LIPIcs 41, pages 504--518. Schloss Dagstuhl -
  LZI, 2015.

\bibitem{BCDGR11}
L.~Brim, J.~Chaloupka, L.~Doyen, R.~Gentilini, and J.-F. Raskin.
\newblock Faster algorithms for mean-payoff games.
\newblock {\em Formal Methods in System Design}, 38(2):97--118, 2011.

\bibitem{DBLP:journals/corr/BruyereFRR14}
V.~Bruy{\`{e}}re, E.~Filiot, M.~Randour, and J.-F. Raskin.
\newblock Expectations or guarantees? {I} want it all! {A} crossroad between
  games and {MDP}s.
\newblock In {\em Proc. of SR}, {EPTCS} 146, pages 1--8, 2014.

\bibitem{bruyere_STACS2014}
V.~Bruy{\`{e}}re, E.~Filiot, M.~Randour, and J.-F. Raskin.
\newblock Meet your expectations with guarantees: Beyond worst-case synthesis
  in quantitative games.
\newblock In {\em Proc. of STACS}, LIPIcs 25, pages 199--213. Schloss Dagstuhl
  - LZI, 2014.

\bibitem{BruyereMR14}
V.~Bruy{\`{e}}re, N.~Meunier, and J.-F. Raskin.
\newblock Secure equilibria in weighted games.
\newblock In {\em Proc. of CSL-LICS}, pages 26:1--26:26. {ACM}, 2014.

\bibitem{CDFR14}
K.~Chatterjee, L.~Doyen, E.~Filiot, and J.-F. Raskin.
\newblock Doomsday equilibria for omega-regular games.
\newblock In {\em Proc. of {VMCAI}}, LNCS 8318, pages 78--97. Springer, 2014.

\bibitem{CDH10}
K.~Chatterjee, L.~Doyen, and T.~A. Henzinger.
\newblock Quantitative languages.
\newblock {\em {ACM} Transactions on Computational Logic}, 11(4), 2010.

\bibitem{DBLP:journals/iandc/Chatterjee0RR15}
K.~Chatterjee, L.~Doyen, M.~Randour, and J.-F. Raskin.
\newblock Looking at mean-payoff and total-payoff through windows.
\newblock {\em Information and Computation}, 242:25--52, 2015.

\bibitem{CH07}
K.~Chatterjee and T.~A. Henzinger.
\newblock Assume-guarantee synthesis.
\newblock In {\em Proc. of {TACAS}}, LNCS 4424, pages 261--275. Springer, 2007.

\bibitem{KHJ06}
K.~Chatterjee, T.~A. Henzinger, and M.~Jurdzi{\'n}ski.
\newblock Games with secure equilibria.
\newblock {\em Theoretical Computer Science}, 365(1):67--82, 2006.

\bibitem{ChatterjeeHP10}
K.~Chatterjee, T.~A. Henzinger, and N.~Piterman.
\newblock Strategy logic.
\newblock {\em Information and Computation}, 208(6):677--693, 2010.

\bibitem{ClarkeE81}
E.~M. Clarke and E.~A. Emerson.
\newblock Design and synthesis of synchronization skeletons using
  branching-time temporal logic.
\newblock In {\em Proc. of Logics of Programs}, LNCS 131, pages 52--71.
  Springer, 1981.

\bibitem{DBLP:conf/lics/ClementeR15}
L.~Clemente and J.-F. Raskin.
\newblock Multidimensional beyond worst-case and almost-sure problems for
  mean-payoff objectives.
\newblock In {\em Proc. of LICS}, pages 257--268. {IEEE}, 2015.

\bibitem{colcombet12}
T.~Colcombet.
\newblock Forms of determinism for automata.
\newblock In {\em Proc. of STACS}, LIPIcs 14, pages 1--23. Schloss Dagstuhl -
  LZI, 2012.

\bibitem{df11}
W.~Damm and B.~Finkbeiner.
\newblock Does it pay to extend the perimeter of a world model?
\newblock In {\em Proc. of {FM}}, {LNCS} 6664, pages 12--26. Springer, 2011.

\bibitem{EM79}
A.~Ehrenfeucht and J.~Mycielski.
\newblock Positional strategies for mean payoff games.
\newblock {\em International Journal of Game Theory}, 8:109--113, 1979.

\bibitem{Faella09}
M.~Faella.
\newblock Admissible strategies in infinite games over graphs.
\newblock In {\em Proc. of {MFCS}}, LNCS 5734, pages 307--318. Springer, 2009.

\bibitem{filar1997}
J.~Filar and K.~Vrieze.
\newblock {\em Competitive {M}arkov decision processes}.
\newblock Springer, 1997.

\bibitem{FKR95}
J.~A. Filar, D.~Krass, and K.~W. Ross.
\newblock Percentile performance criteria for limiting average {M}arkov
  decision processes.
\newblock {\em Transactions on Automatic Control}, 40:2--10, 1995.

\bibitem{fgr10}
E.~Filiot, T.~Le~Gall, and J.-F. Raskin.
\newblock Iterated regret minimization in game graphs.
\newblock In {\em Proc. of {MFCS}}, {LNCS} 6281, pages 342--354. Springer,
  2010.

\bibitem{FismanKL10}
D.~Fisman, O.~Kupferman, and Y.~Lustig.
\newblock Rational synthesis.
\newblock In {\em Proc. of {TACAS}}, LNCS 6015, pages 190--204. Springer, 2010.

\bibitem{hp12}
J.~Y. Halpern and R.~Pass.
\newblock Iterated regret minimization: A new solution concept.
\newblock {\em Games and Economic Behavior}, 74(1):184--207, 2012.

\bibitem{hp06}
T.~A. Henzinger and N.~Piterman.
\newblock Solving games without determinization.
\newblock In {\em Proc. of {CSL}}, LNCS 4207, pages 395--410. Springer, 2006.

\bibitem{hpr15}
P.~Hunter, G.~A. P{\'{e}}rez, and J.-F. Raskin.
\newblock Reactive synthesis without regret.
\newblock In {\em Proc. of CONCUR}, LIPIcs 42, pages 114--127. Schloss Dagstuhl
  - LZI, 2015.

\bibitem{DBLP:dblp_journals/mst/KhachiyanBBEGRZ08}
L.~Khachiyan, E.~Boros, K.~Borys, K.~Elbassioni, V.~Gurvich, G.~Rudolf, and
  J.~Zhao.
\newblock On short paths interdiction problems: Total and node-wise limited
  interdiction.
\newblock {\em Theory of Computing Systems}, 43:204--233, 2008.

\bibitem{KupfermanPV14}
O.~Kupferman, G.~Perelli, and M.~Y. Vardi.
\newblock Synthesis with rational environments.
\newblock In {\em Proc. of {EUMAS}}, LNCS 8953, pages 219--235. Springer, 2014.

\bibitem{mannor_ICML2011}
S.~Mannor and J.~Tsitsiklis.
\newblock Mean-variance optimization in {M}arkov decision processes.
\newblock In {\em Proc. of ICML}, pages 177--184. Omnipress, 2011.

\bibitem{MogaveroMV10}
F.~Mogavero, A.~Murano, and M.~Y. Vardi.
\newblock Reasoning about strategies.
\newblock In {\em Proc. of {FSTTCS}}, LIPIcs 8, pages 133--144. Schloss
  Dagstuhl - LZI, 2010.

\bibitem{nash50}
J.~Nash.
\newblock Equilibrium points in $n$-person games.
\newblock {\em PNAS}, 36:48--49, 1950.

\bibitem{Puterman94}
M.~Puterman.
\newblock {\em Markov decision processes: discrete stochastic dynamic
  programming}.
\newblock Wiley, 1st edition, 1994.

\bibitem{QueilleS82}
J.-P. Queille and J.~Sifakis.
\newblock Specification and verification of concurrent systems in {CESAR}.
\newblock In {\em Proc. of International Symposium on Programming}, LNCS 137,
  pages 337--351. Springer, 1982.

\bibitem{RRS-cav15}
M.~Randour, J.-F. Raskin, and O.~Sankur.
\newblock Percentile queries in multi-dimensional {M}arkov decision processes.
\newblock In {\em Proc. of CAV}, LNCS 9206, pages 123--139. Springer, 2015.

\bibitem{DBLP:conf/vmcai/RandourRS15}
M.~Randour, J.-F. Raskin, and O.~Sankur.
\newblock Variations on the stochastic shortest path problem.
\newblock In {\em Proc. of VMCAI}, LNCS 8931, pages 1--18. Springer, 2015.

\bibitem{Ummels06}
M.~Ummels.
\newblock Rational behaviour and strategy construction in infinite multiplayer
  games.
\newblock In {\em Proc. of {FSTTCS}}, LNCS 4337, pages 212--223. Springer,
  2006.

\bibitem{Vardi-focs85}
M.~Y. Vardi.
\newblock Automatic verification of probabilistic concurrent finite state
  programs.
\newblock In {\em Proc. of FOCS}, pages 327--338. IEEE, 1985.

\bibitem{WL99}
C.~Wu and Y.~Lin.
\newblock Minimizing risk models in {M}arkov decision processes with policies
  depending on target values.
\newblock {\em Journal of Mathematical Analysis and Applications},
  231(1):47--67, 1999.

\bibitem{zjbp08}
M.~Zinkevich, M.~Johanson, M.~Bowling, and C.~Piccione.
\newblock Regret minimization in games with incomplete information.
\newblock In {\em Proc. of NIPS}, pages 905--912, 2008.

\bibitem{ZP96}
U.~Zwick and M.~Paterson.
\newblock The complexity of mean payoff games on graphs.
\newblock {\em Theoretical Computer Science}, 158(1):343--359, 1996.

\end{thebibliography}

\end{document}